\documentclass[12pt]{iopart}
\usepackage{graphicx}
\usepackage{psfig,epsfig}
\usepackage{dcolumn}% Align table columns on decimal point
\usepackage{bm}% bold math
\def\bs{\boldsymbol}   
      
\def\bec{\begin{center}}    \def\eec{\end{center}}
\def\bct{\begin{center}}    \def\ect{\end{center}}  
    
\def\bmp{\begin{minipage}}    \def\emp{\end{minipage}}
\def\beq{\begin{equation}}    \def\eeq{\end{equation}}
\def\bea{\begin{eqnarray}}    \def\eea{\end{eqnarray}}
\def\bes{\begin{eqnarray*}}    \def\ees{\end{eqnarray*}}
\def\bpm{\begin{pmatrix}} \def\epm{\end{pmatrix}}
\def\ben{\begin{enumerate}} \def\een{\end{enumerate}}
\def\btb{\begin{tabular}} \def\etb{\end{tabular}}
\def\btbb{\begin{tabbing}} \def\etbb{\end{tabbing}}
 \def\ie{{\it i.e.\;}} \def\cf{{\it cf.\;}} \def\eg{{\it e.g.\;}}
     
     \def\pt{p_{\rm t}}

 \def\ben{\begin{enumerate}} \def\een{\end{enumerate}}
   \def\btm{\begin{itemize}} \def\etm{\end{itemize}}

   \def\g{\right}  
\def\f{\left}

\begin{document}

\title[]{An Empirical Study of Boosted Neural Network
for Particle Classification in High Energy Collisions}

\author{Meiling Yu, Mingmei Xu and Lianshou Liu}

\address{Institute of Particle Physics, Huazhong Normal University,
Wuhan 430079, China}
\ead{yuml@iopp.ccnu.edu.cn, liuls@iopp.ccnu.edu.cn}

\def\EE{{\rm e}$^+${\rm e}$^-$}

\begin{abstract}
The possible application of boosted neural network to particle
classification in high energy physics is discussed. A
two-dimensional toy model, where the boundary between signal and
background is irregular but not overlapping, is constructed to
show how boosting technique works with neural network. It is found
that boosted neural network not only decreases the error rate of
classification significantly but also increases the efficiency and
signal-background ratio. Besides, boosted neural network can avoid
the disadvantage aspects of single neural network design. The
boosted neural network is also applied to the classification of
quark- and gluon- jet samples from Monte Carlo \EE collisions,
where the two samples show significant overlapping. The
performance of boosting technique for the two different boundary
cases
--- with and without overlapping is discussed.
% and a
%possible method for raising the reconstruction efficiency of
%multi-strange baryon identification is suggested.
\end{abstract}

\vspace{2pc}
\quad particle classification \quad error rate \quad efficiency

\maketitle

\section{Introduction}

Particle identification is very important in the physics of high
energy collisions, and especially, of relativistic heavy-ion
collisions. For example, the identification of multi-strange
baryons with high efficiency and high signal-background ratio is
essential for the study of elliptic flow. The popular method used
in the identification of multi-strange baryons is based on
topological reconstruction. In this method the signals are
extracted from a large amount of combinatoric background by
cutting on certain parameters. This method is reliable, but the
reconstruction efficiency is low. It is about $2\%$ -- $10\%$ in
central and 7\% -- 25\% in peripheral collisions for
$\Xi$~\cite{thesis} and even much lower for $\Omega$. In addition,
to optimize the cuts in a multi-dimensional space by trial and
error can be very tedious. Therefore, a method for raising the
reconstruction efficiency of the identification of this kind of
particles is highly sought.

An alternative method, the artificial neural network has been
introduced into high energy physics in 1988~\cite{Denby1988} and
has been widely used in particle classification such as quark- and
gluon- jets separation~\cite{qg1}\cite{qg2}, photon hadron
discrimination~\cite{photonhadron}, top quark and Higgs
search~\cite{topquark}\cite{higgs1}\cite{higgs2}. Most of the
applications proved that neural network method is superior to
traditional cut method or statistical likelihood method. The
success of neural network method is mainly due to its nonlinear
property, which enables it to explore many hypotheses
simultaneously and consider the correlations between all
variables. Nonetheless, there are also disadvantages when
implementing this method, \eg the final result is more or less
influenced by the design of the architecture and the
initialization of the weight matrices. The effects of these
factors are hard to follow and there is no universal instruction
to help choosing the best parameters.

Boosting is a kind of adaptive reweighting and combining approach
that combines several weak learners into a strong one. It can be
applied to unstable classifiers such as decision trees and neural
networks. Ref~\cite{BDT} claims that boosted decision tree
performs better than artificial neural network in the
neutrino-oscillation search in MiniBooNE experiment. Naturally,
one will ask the question ``how about boosted neural networks?''
Although plenty of studies on UCI machine learning database show
that the performance of boosted neural network is better than that
of single neural network and boosted decision
tree~\cite{boostnn1}\cite{boostnn2}\cite{boostnn3}, so far we have
not seen the application of boosted neural network in high energy
physics.

In this article, we will first discuss briefly the reason why the
efficiency of topological reconstruction method in the
identification of multi-strange baryons is low. Then in Section 3
a brief introduction to neural network and boosting technique will
be given. In Section 4 we will show how boosting works with neural
network in the case of a two-dimensional toy model, where the
boundary of signal and background is irregular but not overlapping. In
Section 5 we will apply boosted neural network to Monte Carlo
quark- and gluon- jets classification, where the two data sets are
overlapping. In the last section we will discuss the performance
of boosted neural network at two different
boundary cases
--- with and without overlapping and a possible
method for raising the efficiency of multi-strange baryon
identification is proposed.

\section{The efficiency of topological reconstruction method}

\begin{table}
\caption {\label{decaytable}Decay parameters of multi-strange baryons.}
{\begin{center} {\footnotesize
\begin{tabular}{ccccc}
\br
Particle & decay mode & fraction (\%) & c$\tau$ (cm) &  Mass (MeV/$c^{2})$  \\
\mr
$\Lambda^{0}$ & p$\pi^{-}$ & 63.9$\pm$0.5 & 7.89 &1115.684$\pm$0.006  \\
%\hline
$\Xi^{-}$ & $\Lambda^{0}\pi{-}$ &99.887$\pm$0.035 & 4.91 &1321.32$\pm$0.13  \\
%\hline
$\Omega^{-}$ & $\Lambda^{0}K^{-}$ & 67.8$\pm$0.7 & 2.46 &  1672.45$\pm$0.29  \\
\br
\end{tabular}
}
\end{center}
}
\end{table}

Traditionally, the strange particles with two-body decay,
$\Xi^{-}$, $\Omega^{-}$ and $\Lambda$, are detected through their
decay topology. The properties of these decays are summarized in
Table~\ref{decaytable}~\cite{pdg}.

Let us take $\Xi^{-}$ search as an example. The primary decay   %%% \dot{}
channel $\Xi^{-}\rightarrow\Lambda\pi^{-}$ has a 99.9$\%$
branching ratio. The daughter particle $\Lambda$ further decays
into $\Lambda\rightarrow p\pi^{-}$ with a 63.9$\%$ branching
ratio. $\Xi^{-}$'s are found by tracing the decay topology
backwards. First, a neutral decay vertex is found by identifying
the crossing points of positive and negative particles' tracks.
Kinematic information about the tracks are used to determine the
trajectory of the parent neutral particle. The neutral particle is
then intersected with other negative tracks to obtain candidate
$\Xi^{-}$ decay vertices. A schematic diagram of a $\Xi^{-}$ decay
is given in figure 1.

\begin{figure}
\centering
\includegraphics[width=0.5\textwidth]{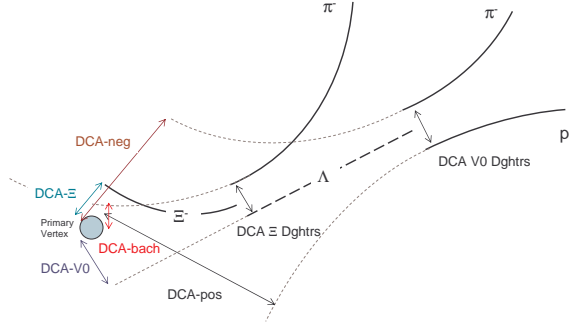}
%\makeatletter\def\@captype{figure}\makeatother
\caption{\label{Fig. 1} Schematic representation of a $\Xi^{-}$
decay with distance of closest approach (DCA) parameters. }
\end{figure}

In each Au-Au collision event at RHIC energy ($\sqrt {s_{\rm
NN}}=200$ GeV) up to several thousand particles are produced. The
finite momentum resolution of the TPC causes the primary tracks to
not point back exactly to the primary vertex. As a result, these
tracks may randomly cross with other primary tracks and form fake
secondary vertices. Indeed, in the quagmire of particle tracks,
the vertices can be quite easily misidentified, leading to a large
combinatoric background. To reduce this background, basic cuts are
applied during the event reconstruction chain.

To determine if two tracks are originated from the same vertex, a
cut is placed on their distance of closest approach (DCA). This
cut reduces the random background by a large amount, but is
insufficient to guarantee a good identification of the parent
particle. Other cuts are necessary because of the following
reasons:
\begin{table}
\caption {\label{cuts}Track quality and kinematic cuts for $\Xi^{-}$
reconstruction.}  {\footnotesize
\begin{tabular}{lll}
\br
Track Selection Criteria & loose cut & tight cut \\
\mr V0 Vertex Cuts: &&\\
\mr
proton TPC hits & $\geq$15 & $\geq$20   \\
pion TPC hits & $\geq$15 & $\geq$17   \\
pion and proton PID (for $0.0<\pt<2.0$ GeV/$c$) &$\leq3\sigma$&$\leq3\sigma$\\
Track proton DCA to primary vertex(cm) & $\geq$0.5 & $\geq$0.5  \\
\qquad(for $0.0<\pt<2.0$ GeV/$c$)  & & \\
Track pion DCA to primary vertex (cm)& $\geq$2.0 & $\geq$2.0  \\
%%%%%%%%%%%%  dca V0 daughters (cm)& $\leq$0.7&$\leq$0.65\\
DCA between V0 daughters (cm)& $\leq$0.7&$\leq$0.65\\
V0 DCA to primary vertex (cm)& $\geq 0$ and $\leq$0.7 & $\geq 0.35$ and $\leq 0.7$  \\
V0 decay length from primary vertex ($r1$)(cm)&$\geq$5.0&$\geq$5.0\\
\mr $\Xi$ Vertex Cuts: && \\
\mr V0 mass window (Mev/$c^{2}$)&$\pm7$&$\pm4 $\\
bachelor $\pi^{-}$ TPC hits &$\geq$10&$\geq$17\\
%%%%%% dca bachelor $\pi^{-}$ to primary vertex (cm)&$\geq$0&$\geq$0.5\\
bachelor $\pi^{-}$ DCA to primary vertex (cm)&$\geq$0&$\geq$0.5\\
bachelor PID (for $0.0<\pt<2.0$ GeV/$c$) &$\leq3\sigma$&$\leq3\sigma$\\
bachelor $\pt$ (GeV/$c$)&$\geq0.075$&$\geq0.075$\\
%%%%%%% dca $\Xi$ daughters (cm)&$\leq$0.7&$\leq$0.65\\
DCA between $\Xi$ daughters (cm)&$\leq$0.7&$\leq$0.65\\
angle between $\Xi$'s momentum and decay vertex vector&$\leq
\arccos(0.9)$&$\leq \arccos(0.9)$\\
%%%%%%% $\Xi$ dca to primary vertex (cm)&$\leq$1.0&$\leq$0.5\\
$\Xi$ DCA to primary vertex (cm)&$\leq$1.0&$\leq$0.5\\
$\Xi$ decay length from primary vertex ($r2$)(cm)&$\geq 2.0$ and $\leq r1$&$\geq 2.5$ and $\leq r1$\\
\br
\end{tabular}
}
\end{table}
\begin{itemize}
\item Due to high density of tracks near the primary vertex, it is
easy to form many fake track crossings. This leads to a larger
combinatoric background as one gets closer to the primary vertex.
The decay distance distribution has an exponential fall-off to
zero, so cuts used on these distances for the candidate $\Xi^{-}$
and daughter $\Lambda$ are greater than 2cm and 5cm, respectively.
The decay distances are measured from the primary vertex.

\item The candidate parents, \ie candidate $\Xi^{-}$'s, point back to the primary
vertex since %%% heavier particles are produced near hear.
they are produced at this vertex.

\item The
daughter tracks do not point back to the primary vertex to ensure
they are not primary tracks.

\item A cut on the calculated mass of
the daughter neutral particle is added to increase the likelihood
that the parent particle did indeed decay into a $\Lambda$
($\bar{\Lambda}$) plus a charged track.
\end{itemize}

\bigskip
Typical cuts used for $\Xi^{-}$ are listed in Table~\ref{cuts}. For track
pairs passed all the cuts the invariant mass for a decay vertex is
calculated by the kinematic information of its daughters:
$$
m=\sqrt{\f(\sqrt{m_{1}^{2}+P_{1}^{2}}+\sqrt{m_{2}^{2}+P_{2}^{2}}\g)^{2}-P^{2}},
$$
where subscripts 1 and 2 represent the two daughter particles from
a decay vertex. This equation is used twice since there are two
decay steps associated with a $\Xi^{-}$ particle. For $\Lambda$
reconstruction one of them is $p$
 and the other is $\pi^{-}$. For $\Xi^{-}$ reconstruction one of them
is $\Lambda$ and the other is bachelor $\pi^{-}$. $P=P_{1}+P_{2}$
is the parent momentum with $P_{1}$ and $P_{2}$ representing the
daughters' momentum. With all the topological cuts used, clear
signal peak is observed in invariant mass distribution.

A tighter cut will make the peak becomes more significant.
However, the extremely evil cut results in a high
signal-background ratio but at the same time it reduces the signal
yield greatly, causing the reconstruction efficiency to be very
low.

\section{A brief description of neural network and boosting technique}

The neural network approach is nothing but functional fitting to
data. In classification cases, one wants to construct a mapping
$F$ between a set of observable quantities $x_i$ ($i=1,\dots,s$)
and category variable $Y$ by fitting $F$ to a set of $M$ known
``training'' samples $(x_i^{(p)},
y_k^{(p)},\,i=1,\dots,s;\,k=1,\dots,v) (p=1,\dots, M,\, y_k\in
Y)$. Once the parameters in $F$ are fixed, one then uses this
parametrization to interpolate and find the category of ``test''
samples not included in the ``training'' set. Obviously, the
performance of the network on the test set estimates the
generalization ability of the fitting. In the present work we use
the multilayer perceptron program developed in ROOT version
4.00/04~\cite{root}. The function $F$ is an expansion of sigmoidal
function in a feed forward network structure since there is a
theorem~\cite{theorem1} saying that a linear combination of
sigmoids can approximate any continuous function.

A typical three-layer neural network is sketched in
figure~\ref{nnstructure}. It consists of an input layer, a hidden
layer and an output layer, with various number of nodes (also
called neurons) in each layer. In the following, we will use the
notation [$s$-$u$-$v$] to denote a neural network with $s$ input
nodes, $u$ hideen nodes and $v$ output nodes. There are weights
connecting the nodes from any two adjacent layers and each node in
the hidden and the output layers has a threshold. The output of
the $k$th node in the output layer ($k=1,\dots,v$) is $$
O_k(x_1,x_2,\cdots,x_s)=f\left(\sum_{j=1}^u
w'_{jk}x'_j-\theta'_k\right)= f\left(\sum_{j=1}^u
w'_{jk}f\left(\sum_{i=1}^s
w_{ij}x_i-\theta_j\right)-\theta'_k\right), $$ where $\{x_i\}$ are
the input parameters, $\{w_{ij}\}$ are weights between the $i$th
node in the input layer and the $j$th node in the hidden layer,
$\{\theta_j\}$ are thresholds of each node in the hidden layer,
$\{w'_{jk}\}$ are weights between the $j$th node in the hidden
layer and the $k$th node in the output layer, $\{\theta'_k\}$ are
thresholds of each node in the output layer. $f(x)=1/(1+e^{-x})$
is the sigmoid transfer function.

The goal of adjusting the parameters, or training the neural
network, is to minimize the fitting error. The mean square error
$E$ averaged over the training samples is defined as $$ E={1\over
{2M}}\sum_{p=1}^M\sum_{k=1}^v(O_k^{(p)}-y_k^{(p)})^2,$$ where
$O_k$ is the output of the $k$th node of the neural network, $y_k$
is the training target, $M$ is the number of samples in the
training set. In binary case, the output layer has only one node,
 $v=1$, with $y_1=0$ for background and $y_1=1$ for signal. There are
several algorithms for error minimization and weight updating,
which are implemented in ROOT as options. The initial weights are
random numbers in the range $(-0.5, 0.5)$.

\begin{figure}
\begin{center}
\includegraphics[width=2.0in]{./Figures/nnstructure.epsi} \hskip0.5cm
\includegraphics[width=2.0in]{./Figures/toydis.epsi}
\begin{minipage}[t]{150pt}
\caption{\label{nnstructure} A sketch of multilayer perceptron.}
\end{minipage} %\hskip1.5cm
\begin{minipage}[t]{250pt}
\caption{\label{toydis} The distribution of signal and background
from the two-dimensional toy model, where crosses are signals and
circles are backgrounds.}
\end{minipage}
\end{center}
\end{figure}

Boosting is a technique to construct a committee of weak learners
that lowers the  error rate in classification. It is first
developped by Schapire~\cite{Schapire90} and the theoretical study
followed shows that  given a significant number of weak learners,
the boosting algorithm can decrease the error rate on the training
set and convert the ensemble of weak learners to a strong learner
whose error rate on the ensemble is arbitrarily
low~\cite{Freund95}\cite{Friedman}. In the binary classification
case, one only needs to construct weak learners with error rate be
slightly better than random guessing (0.5). There are a number of
variations on basic boosting. The most popular one, AdaBoost,
allows the designer to continue adding weak learners until the
desired low training error has been achieved. In AdaBoost each
training sample receives a distribution $D(p)$ that determines its
probability of being selected in a training set for individual
component classfier. The distribution $D(p)$ is determined in the
following way --- if a training sample is accurately classified,
then its chance of being used again in a subsequent component
classifier is reduced; conversely, if the pattern is not
accurately classified, then its chance of being used again is
raised. Thus AdaBoost ``focuses'' the component classifier on more
informative or  more ``difficult'' samples.

The AdaBoost procedure for neural network in binary case is as follows:

{\bf Input:} sequence of $M$ examples $({\bs x}^1,y^1),\dots,
({\bs x}^M,y^M)$ with labels $y^p\in\{0,1\}$, here ${\bs
x}={\{x_i\}}\,(i=1,\dots,s)$.

{\bf Init:} $D_0(p)=1/M$ for all $p$.

{\bf Repeat ($t=0,\dots,T$):}
\begin{itemize}
\item[1.] Train neural network with respect to distribution $D_t$ and obtain
hypothesis $h_t:X \rightarrow Y$,
\item[2.] calculate the weighted error of $h_t:$
$\epsilon_t=\sum_{p:h_t({\bs x}^p)\ne y^p}D_t(p)$ and abort loop
if $\epsilon_t>{1\over 2}$,
\item[3.] set $\alpha_t={1\over
2}\ln{{1-\epsilon_t}\over\epsilon_t}$,
\item[4.] update distribution $D_t$
$$ D_{t+1}(p)={D_t(p)\over Z_t}\times\left\{\begin{array}{ll}
e^{\alpha_t}\quad & \mbox{ if } h_t({\bs x}^p)\ne y^p \mbox{
(incorrectly classified),}\\ e^{-\alpha_t}\quad & \mbox{ if }
h_t({\bs x}^p)=y^p \mbox{ (correctly
classified),}\end{array}\right.$$ where $Z_t$ is a normalizing
constant.
\end{itemize}

{\bf Output:} final hypothesis: $h_{\rm final}({\bs
x}^p)=\arg\max\limits_{y}\sum\limits_t\alpha_t h_t({\bs x}^p)$.

In binary case the final hypothesis can be restated as $$ h_{\rm
final}({\bs x}^p)=\left\{\begin{array}{l} 1 \quad\mbox{ if }
\sum_{t=1}^T\alpha_th_t({\bs x}^p)\ge {1\over
2}\sum_{t=1}^T\alpha_t, \\ 0 \quad\mbox{
otherwise.}\end{array}\right.$$

\section{A two-dimensional toy model}

In high energy physics, the separation of signal from background
is a typical binary case. Assume a data set with $n_s$ signals and
$n_b$ backgrounds. Neural network is applied to it and the result
can be denoted by the following quantities:
\begin{itemize}
\item[] $n_{cs}:$ the number of signals correctly classified
\item[] $n_{ws}:$ the number of signals incorrectly classified
\item[] $n_{cb}:$ the number of backgrounds correctly classified
\item[] $n_{wb}:$ the number of backgrounds incorrectly classified
\end{itemize}
The classification ability of the neural network can be judged by
the error rate, classification efficiency and signal-background
ratio, with the definitions: $$ \mbox{error
rate}={{n_{ws}+n_{wb}}\over{n_s+n_b}}, \qquad
\mbox{efficiency}={n_{cs}\over n_{s}}, \qquad \mbox{S-B ratio}
={n_{cs}\over n_{wb}}.$$ According to the above definition, even
though boosting is capable in decreasing the error rate of the
training set, this does not imply that the classification
efficiency and signal-background ratio could be increased. In
physics, what is most sought for is high efficiency and high
signal-background ratio.

In this section we construct a two-dimensional toy model to show
how the boosting algorithm works with neural network for the case
of non-overlapping boundary between signal and background, \cf
figure~\ref{toydis}. In the figure, crosses are signals and circles
are backgrounds. The boundary between them is irregular but does
not overlap.

\begin{figure}
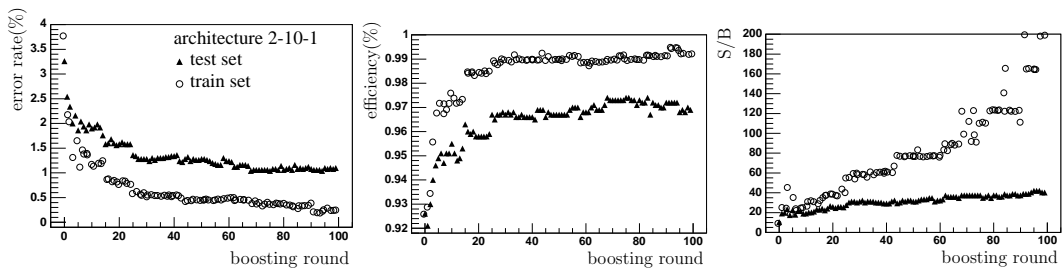

\begin{center}
\includegraphics[width=1.8in]{./Figures/toyer.epsi}
\includegraphics[width=1.8in]{./Figures/toyef.epsi}
\includegraphics[width=1.8in]{./Figures/toysb.epsi}
\caption{\label{toy} The error rate, classification efficiency and
signal-background ratio of the classification by boosted neural
network for both training and test sets under the network
architecture 2-10-1.}
\end{center}
\end{figure}

To train the neural network we require similar signal- and
background- sample-density in the phase space of the training set.
Thus 1000 signals and 4000 backgrounds are used to form the
training set and another set with the same amount of samples are
used as test set. The inputs of the network are the coordinates of
the points in X-Y plane. Firstly, we choose a three-layer network
[2-10-1] with two input nodes, ten hidden nodes and one output
node. All the other network parameters take the default values
from ROOT. In Figure~\ref{toy} are shown the error rate, the
classification efficiency and the signal-background ratio of the
classification for both training and test sets during being
boosted one hundred rounds. We see that with respect to the
boosting round, the error rates of the training and test sets
decreas sharply while the classification efficiency increases at
the first few rounds then slightly oscillates afterwards. The
signal-background ratio also increases. Obviously, a boosted
neural network can distinguish signal from background better than
single neural network with the same network parameters.

\begin{table}
\caption{\label{toytable}The error rate, classification efficiency
and signal-background ratio  of the classification by single and
boosted neural networks with different network architectures and
different initial weight matrices in test set for the two-dimensional
toy model. w1--w7 indicate seven different initial
weight matrices.}
\begin{indented}
\lineup
\item[]\begin{tabular}{@{}*{10}{l}}
\br
architecture&&&w1&w2&w3&w4&w5&w6&w7\\
\mr
                             &     &error rate (\%)&5.94&4.93&4.46&4.30&5.90&4.40&5.96\\
                             &single&efficiency (\%)&89.1&78.6&92.9&93.7&89.6&93.3&89.0\\
 \raisebox{-2.3ex}[0pt]{\ \quad 2-5-1} &     &S-B ratio&4.74&3.42&6.11&6.16&4.69&6.10&4.73\\
                           &     &error rate (\%)&1.04&0.60&0.52&0.94&0.94&0.94&0.98\\
                            &boosted&efficiency (\%)&96.9&97.4&99.4&96.5&98.4&97.4&97.2\\
                             &     &S-B ratio&46.1&34.8&49.7&80.4&31.7&46.4&46.3\\
\mr
                              &      &error rate (\%)&2.94&4.10&4.68&3.26&2.66&3.08&3.06\\
                              &single&efficiency (\%)&94.6&92.4&89.7&92.6&94.2&93.6&94.5\\
\raisebox{-2.3ex}[0pt]{\ \quad 2-10-1}&      &S-B ratio&10.2&7.16&6.85&10.4&12.6&10.4&9.64\\
                              &      &error rate (\%)&0.88&1.10&0.94&1.04&1.02&0.98&0.88\\
                              &boosted&efficiency (\%)&97.3&96.7&97.0&97.4&98.2&97.5&97.5\\
                              &      &S-B ratio&57.3&44.0&57.1&37.5&29.8&40.6&51.3\\
\mr
                               &      &error rate (\%)&2.32&1.48&1.6&2.38&1.46&1.54&2.38\\
                               &single&efficiency (\%)&93.0&95.6&95.6&92.4&95.8&95.6&94.4\\
\raisebox{-2.3ex}[0pt]{\ \quad 2-5-4-1} &     &S-B ratio&20.2&31.9&26.6&21.5&30.9&28.9&15.0\\
                               &      &error rate (\%)&0.88&0.64&0.94&0.60&0.88&0.76&0.70\\
                               &boosted&efficiency (\%)&97.3&98.6&97.0&98.6&97.5&97.6&98.8\\
                               &     &S-B ratio&57.2&54.7&57.1&61.6&51.3&75.1&43.0\\
\br
\end{tabular}
\end{indented}
\end{table}

One of the disadvantages of neural network is that its performance
depends on network architecture and initial weight matrices. To
study the dependency, we vary the hidden nodes of the network and
use different weight matrices. The behavior of the networks in
terms of error rate, classification efficiency and
signal-background ratio are listed in Table~\ref{toytable}.

It can be seen from the table that, in average, more complicated
single network architecture, \eg [2-5-4-1], gives out lower error
rate, higher efficiency and higher signal-background ratio than
the other two architectures, and that for different initial weight
matrices the performance of the network varies even for the same
architecture. After boosting, not only the error rate decreases
but also the classification efficiency and signal-background ratio
increase in comparison to those of the single neural network. In
addition, the dependency of the network performance on different
network architectures and initial weight matrices vanishes.
Therefore, boosted simple neural network with arbitrary initial
weight matrices has comparable ability as a single neural network
with a complicated structure and fine-tuned parameters.

\section{Classification of quark- and gluon- jets}
\def\pt{p_{\rm t}}  \def\Evis{E_{\rm vis}}
The Monte Carlo JetSet7.4 is used to generate $\rm e^+\rm e^-$
collision events at 91.2 GeV. The quark- and gluon- jet samples
are obtained through the following procedure: (1) Force events
into 3-jet ones both at parton and at hadron levels by using the
Durham jet algorithm~\cite{Durham}. (2) Select the planar 3-jet
events by requiring the sum of the three angles between two
adjacent jets to be greater than $358^{\circ}$ at hadron level
(this condition is automatically satisfied at parton level). (3)
Apply the angular cut method~\cite{yu} to the hadronic 3-jet
event, {\it i.e.} the three angles between two adjacent jets are
ordered and the jet opposite to the largest angle is supposed to
be a gluon jet and the jet opposite to the smallest angle is the
more energetic quark jet. We require the difference between the
largest angle and the middle one to be greater than an angular cut
$20^{\circ}$ and the more energetic quark jet is rejected in an
event. (4) Match the hadronic quark- and gluon- jets with the
corresponding parton level jets. Four variables are chosen to
describe a jet, \ie the multiplicity $n$ inside jet, the
transverse momentum $\pt$ of jet, the included angle $\theta$
opposite to the jet and the jet energy $\Evis$.

Usually, the quark and gluon jet samples are hard to be
distinguished because of the large overlapping of these two sets.
Using the above procedure, the quark- and gluon- jet samples are
well selected from the generated raw samples and the overlapping
in some phase space is decreased, so that we can study how the
network works and compare it with the simple cut method.

In Figure~\ref{qgboundary} is shown the mixing region of quark-
and gluon- jets in $\Evis$ and $\theta$ space. Compared with the
simple two-dimensional toy model, the boundary is unclear and the
two sets overlap each other strongly. If we apply a $\theta$ cut,
\eg setting $\theta\geq 2.675$ to be gluon jet and $\theta<2.675$
to be quark jet, the efficiency for both quark- and gluon- jets
are greater than $90\%$ and the error rate is $4.05\%$

\begin{figure}
\begin{center}
\includegraphics[width=2.2in]{./Figures/qgboundary.epsi} \hskip0.5cm
\includegraphics[width=2.2in]{./Figures/qger.epsi}
\begin{minipage}[t]{210pt}
\caption{\label{qgboundary} The distribution of quark- and gluon-
jets near their boundary. Crosses represent quark jets and circles
represent gluon jets. }
\end{minipage}
\begin{minipage}[t]{210pt}
\caption{\label{qger} The error rate of the classification of
quark and gluon jet samples by boosted neural network both for
training set and for test set under network architecture 4-5-1.}
\end{minipage}
\end{center}
\end{figure}

Next, we take 2500 quark jets and 2500 gluon jets as training set
and another 5000 as test set. From the above two-dimensional toy
model we know that boosting different network architectures will
result in similar performance, so we choose a very simple network
architecture [4-5-1] to save the computer time and ensure network
generalization ability.

In figure~\ref{qger} are plotted the error rates of the network in
the training and test sets versus the boosting round.
%It can be
%seen from the figure that the error rate of training set decreases
%with respect to the increasing of the boosting round, just as what
%we see in the two-dimensional toy model, while the behavior of the
%test set is not as expected. It increases in the first round, then
%decrease slightly and oscillates, keeping higher than the error
%rate of the first round.
It can be seen from the figure that the error rates increase in
the first boosting round then decrease afterwards. For the
training set the error rate decreases to zero while for the test
set, the error rate decreases slightly and oscillates, keeping
higher than that of the first round. To check the results we tried
a more complicated network architecture [4-50-1] to do the
boosting. The similar trends are obtained. The comparison of
single neural network, boosted network and simple cut method with
respect to error rate, efficiency and signal-background ratio are
shown in Table~\ref{qgtable}. It can be seen that, although the
boosted neural network does not work well for the present case,
the single neural network still presents lower error rate, higher
efficiency and higher signal-background ratio in comparison with
the simple cut method.

\begin{table}
\begin{center}
\caption{\label{qgtable}The error rate, classification efficiency
and signal-background ratio of the single neural network, the
first boosting round and the boosted neural network with different
network architectures and different initial weight matrices for
test set in the case of quark- and gluon-jets classification. }
{\small
\begin{tabular}{@{}*{11}{l}}
\br
&&\centre{3}{error rate (\%)}&\centre{3}{efficiency(\%)}&\centre{3}{S-B ratio}\\
\ns
\multicolumn{2}{c}{Methods}&\crule{3}&\crule{3}&\crule{3}\\
&&w1&w2&w3&w1&w2&w3&w1&w2&w3\\
\mr
     &single&1.42&1.38&1.50&98.80&98.80&98.64&60.2&63.3&60.2\\
4-5-1&1st round&2.22&2.46&2.12&97.96&98.32&98.00&40.8&30.3&43.8\\
     &boosted&1.82&1.60&1.50&98.08&98.32&98.28&57.0&64.7&76.8\\
\mr
      &single&1.36&1.34&1.38&98.80&98.80&98.72&65.0&66.8&66.7\\
4-50-1&1st round&2.72&2.70&2.58&97.68&97.36&97.84&31.3&35.3&32.6\\
      &boosted&1.68&1.72&1.66&98.28&98.04&98.28&59.9&66.2&61.4\\
\mr
%\qquad \quad
\qquad simple&\hskip-0.2cm cut& &4.05& & &97.65& &&18.0&\\
\br
\end{tabular}
}
\end{center}
\end{table}

\section{Discussions}

In this paper we apply the boosting technique to artificial neural
network. A two-dimensional toy model is constructed to show how
boosted neural network works when the boundary between signal and
background is complicated but does not overlap. Then the boosted
neural network is applied to Monte Carlo quark- and gluon- jet
samples of \EE collision, where the two samples strongly mix with
each other.

In both cases the boosting technique drives the error rate of the
training set to zero, while the error rates of the test sets
behave differently. In the case of two-dimensional toy model with
non-overlapping boundary, the error rate of the test set also
decreases dramatically but in the case of quark- and gluon- jet
samples with strong overlap, the error rate of the test set
increases at the first boosting round then decreases slightly. 
We also tried some other event samples, for example, a tighter or looser
angular cut mentioned in section 5 
or cuts on the network input parameter, the included angle $\theta$, 
are applied to quark and gluon
jets to make the classification task easier or harder. We found that once there
are mixing between quark-jet set and gluon-jet set, the performance 
of boosted neural networks is similar as that shown above.
%It means 
So we conclude that boosting technique does 
not improve the performance of single neural networks in the case of
overlapping samples like the quark and gluon jets. 
This is easy to understand since boosting technique is always applied to
unstable classifiers, while we see that from table~\ref{qgtable} the  
outcomes of single neural network are rather stable at different 
network architectures and initial weight matrices. There is no space for
boosting technique to improve the behavior of single neural networks in
this case. 
%not work good in the case of
%overlapping samples.

To summarize, artificial neural network, in general, results in
lower error rate than simple cut method. Boosted neural network
avoids the disadvantage of single neural network and is more
stable and easier to implement. It will lower the error rate,
increase the efficiency and signal-background ratio of
classification in case the boundaries between signal and background
are complicated but separable which could not be easily 
classified by simple cut method.
While for the case with mixed signal and background,  
the boosted neural network does not help improving the classification.

\ack This work is supported in part by the National Science
Foundation of China under project 10375025 and by the Cultivation
Fund of the Key Scientific and Technical Innovation
Project, Ministry of Education of China NO CFKSTIP-704035.

\end{document}